\documentclass[a4paper,fleqn,usenatbib]{mnras}

\usepackage{txfonts}

\usepackage[T1]{fontenc}
\usepackage{ae,aecompl}

\usepackage{psfig}   
\usepackage{graphicx}
\usepackage{amssymb}
\usepackage{subfigure}

\title[Binary destruction in substructure]{A dependence of binary and planetary system destruction on subtle variations in the substructure in young star-forming regions}

\author[R.~J.~Parker]{Richard  J. Parker\thanks{E-mail: R.Parker@sheffield.ac.uk}\thanks{Royal Society Dorothy Hodgkin Fellow} \vspace*{0.1cm} \\
  Department of Physics and Astronomy, The University of Sheffield, Hicks Building, Hounsfield Road, Sheffield, S3 7RH, UK}

\begin{document}

\date{}
                             
\pagerange{\pageref{firstpage}--\pageref{lastpage}} \pubyear{2023}

\maketitle

\label{firstpage}

\begin{abstract}
Simulations of the effects of stellar fly-bys on planetary systems in star-forming regions show a strong dependence on subtle variations in the initial spatial and kinematic substructure of the regions. For similar stellar densities, the more substructured star-forming regions disrupt up to a factor of two more planetary systems. We extend this work to look at the effects of substructure on stellar binary populations. We present $N$-body simulations of substructured, and non-substructured (smooth) star-forming regions in which we place different populations of stellar binaries. We find that for binary populations that are dominated by close ($<$100\,au) systems, a higher proportion are destroyed in substructured regions. However, for wider systems ($>$100\,au), a higher proportion are destroyed in smooth regions. The difference is likely due to the hard-soft, or fast-slow boundary for binary destruction. Hard (fast/close) binaries are more likely to be destroyed in environments with a small velocity dispersion (kinematically substructured regions), whereas soft (slow/wide) binaries are more likely to be destroyed in environments with higher velocity dispersions (non-kinematically substructured regions). Due to the vast range of stellar binary semimajor axes in star-forming regions ($10^{-2} - 10^4$\,au) these differences are small and hence unlikely to be observable. However, planetary systems have a much smaller initial semimajor axis range (likely $\sim$1 -- 100\,au for gas giants) and here the difference in the fraction of companions due to substructure could be observed if the star-forming regions that disrupt planetary systems formed with similar stellar densities.
\end{abstract}


\section{Introduction}

Observations of the Galactic field show that up to 50\,per cent of stellar systems are in fact binary, triple or higher order multiple systems \citep{Duquennoy91,Raghavan10,Ward-Duong15,Tokovinin14}, with slightly lower multiplicity fractions reported for lower-mass primary systems \citep{Bergfors10,Janson12,Ward-Duong15} and slightly higher multiplicities reported for higher-mass primary systems \citep{Sana13,DeRosa14}.

For pre-main sequence, and proto-stellar systems, the multiplicity fraction is much higher, approaching 100\,per cent \citep{Connelley08,Chen13,Pineda15}, which suggests that the vast majority of stellar systems form in multiples, which are then destroyed either by direct dynamical encounters with passing stars \citep{Heggie75,Hills75a,Kroupa95a}, or from dynamical decay \citep{Reipurth12,Reipurth14}.

The efficacy of destruction via dynamical encounters is currently debated, with some authors arguing that a significant number of systems are destroyed due to interactions with passing stars or binaries in star-forming regions \citep{Kroupa95a,Kroupa99,Parker09a,Marks12,Marks14}. Other authors have argued that the amount of dynamical processing required to significantly alter a population of binaries is inconsistent with the observed density and spatial substructure in  many nearby star-forming regions \citep{King12a,King12b,Parker14e}, although recent analysis of \emph{Gaia} data suggest that some regions are dense enough to disrupt significant numbers of binaries \citep{Farias20,Schoettler20,Schoettler22}.

If binaries are processed through dynamical encounters in star-forming regions, then the overall fraction of multiple systems decreases and the semimajor axis and eccentricity distributions are also altered \citep[e.g.][]{Kroupa95a,Marks11,Leigh15,Becker13}. Interestingly, the shape of the mass ratio distribution is not altered by dynamics, even if a significant number of systems have their other orbital parameters altered \citep{Parker13b}.

The dominant factor in whether a population of multiple systems will be dyanmically processed, and the extent to which they are affected, is usually the stellar density of the star-forming region. Most star-forming regions are at least a factor of $\sim$1000 more dense than the Galactic field \citep{Bressert10,Marks12}, and some may be more dense than this at birth \citep{Pfalzner14,Parker17a,Parker22a,Schoettler22}. In addition to the initial, or maximum density, the time spent in a dense region by a population of multiple stars will also affect how many systems are disrupted.

Other, more subtle factors, may also influence whether binary and multiple systems are disrupted. $N$-body, and hybrid $N$-body/hydrodynamical simulations, of star-forming regions now usually adopt spatially and kinematically substructured initial conditions for the initial distributions of stellar systems \citep{Goodwin04a,Parker14b,Dorval16,Sills18,Zwart19,Cloutier21,Torniamenti22}. \citet{Parker11c} showed that spatially and kinematically substructured star-forming regions process more systems than in smoother regions with similar radii. However, these simulations have very different \emph{local} densities, which means that the multiple systems experience a very different encounter history in the substructured regions, compared to the smoother regions.

\citet{Zheng15} performed a similar study to that in \citet{Parker11c}, but with planetary mass objects. They also noted a dependence on the initial substructure, though again their simulations were all set up with similar radii so that the local densities were different. (As an aside, a significant wealth of literature discusses the effects of dynamical encounters on planetary systems in star-forming regions, see e.g. \citet{Laughlin98,Smith01,Adams06,Bonnell01b,Parker12a,Perets12,Craig13,Hao13,Li15,Cai17a,Cai19,Fujii19,FlamminiDotti19,Stock20,Wang22,Ndugu22,Stock22,Carter23,Rickman23}, and many others.)

Recently, \citet{DaffernPowell22} quantified the amount of dynamical processing that could be experienced by planetary systems in substructured star-forming regions, in which they performed a systematic analysis of the effects of spatial and kinematic substructure. To do this, they set the radii of the star-forming regions such that the initial median local densities were the same across all simulations. Therefore, a highly substructured region had a larger radius than the corresponding smooth region, which enables a fair comparison of the effects of substructure on the companion population. \citet{DaffernPowell22} found that the fraction of planets that became liberated from their host star, or free-floating, was almost a factor of two higher in the substructured simulations, even at comparable densities to the smoother simulations.

If changing the amount of substructure in a star-forming region leads to a similar difference in the processed stellar binary populations compared to planetary systems, this could potentially be observable in stellar populations. Furthermore, an investigation is also required to pinpoint why systems with similar stellar densities can process populations to such a different extent. In this paper, we quantify the evolution of multiplicity fractions in simulations with different primoridal populations and compare the results for simulation with and without initial spatial and kinematic substructure (but with comparable stellar densities). The paper is organised as follows. We describe our simulations in Section~\ref{methods}, we present  our results in Section~\ref{results}. We provide a discussion in Section~\ref{discuss}, and we conclude in Section~\ref{conclude}.

\section{Methods}
\label{methods}

In this Section we describe the set-up of the spatially and kinematically substructured, and non-substructured (smooth) star-forming regions,and thier constituet binary populations, which we subsequently evolve as $N$-body simulations.

\subsection{Stellar systems}

The simulations are set up such that there are 1000 stellar \emph{systems} in each star-forming region. This is motivated by the observations of the mass function of star clusters and associations \citep{Lada03}, which has the form 
\begin{equation}
  \frac{dN}{dM} \propto M_{\rm SF}^{-2},
\end{equation}
where $M_{\rm SF}$ is the mass of a star-forming region and this relation is valid for  $M_{\rm SF}$ between $10 - 10^5$\,M$_\odot$. A star-forming region with 1000 systems will have a total mass of $\sim$500\,M$_\odot$, which lies towards the middle of this mass distribution, but is not so massive that it would be considered an unusual star-forming region.

We sample system masses from the probability distribution described in \citet{Maschberger13}, which has the form
\begin{equation}
p(m) \propto \left(\frac{m}{\mu}\right)^{-\alpha}\left(1 + \left(\frac{m}{\mu}\right)^{1 - \alpha}\right)^{-\beta}.
\label{maschberger_imf}
\end{equation}
In Eqn.~\ref{maschberger_imf}  $\mu = 0.2$\,M$_\odot$ is the scale parameter, or `peak' of the IMF \citep{Bastian10,Maschberger13}, $\alpha = 2.3$ is the \citet{Salpeter55} power-law exponent for higher mass stars, and $\beta = 2.0$ describes the slope of the IMF for low-mass objects, taking into account that we have significant numbers of binary systems. We randomly sample this distribution in the mass range 0.01 -- 50\,M$_\odot$ for all simulations, aside from the set from \citet{DaffernPowell22} where there are no brown-dwarfs, so the mass range is 0.08 -- 50\,M$_\odot$. The chosen stellar/substellar mass ranges for each set of simulations are shown in Table~\ref{simulations}.

We determine whether a system is a binary by selecting a random number between zero and unity, and create a binary system if the random number is less than or equal to the binary fraction, defined as
\begin{equation}
f_{\rm bin} = \frac{B}{S + B},  
\end{equation}
where $S$ and $B$ are the respective numbers of single and binary systems. We do not include triples or higher-order multiple systems in our simulations, although such systems are observed to be common in the Galactic field \citep[e.g.][]{Tokovinin08,Tokovinin14,Tokovinin18} and are probably ubiquitous in star formation \citep{Reipurth14,Pineda15}.

In the majority of simulations, we set the binary fraction to $f_{\rm bin} = 0.5$, but we also run a set of simulations where we adopt a version of the \citet{Kroupa95a} `Universal' binary population. In this population, the binary fraction is $f_{\rm bin} = 1$, i.e. all stars are in binaries. However, as we will see, some of the binaries in this population are so wide initially that they are not considered gravitationally bound in the simulations (the widest systems have semimajor axes of order 0.1\,pc, which is similar to the distance between stellar systems in these simulations).

We also run a set of simulations where the initial binary fraction is similar to that observed in the Galactic field. The Galactic field population is characterised by a decreasing binary fraction with decreasing primary mass. Informed by the observations, for this set of simulations we set the brown dwarf binary fraction to $f_{\rm bin} = 0.15$ for systems with primary masses in the range 0.01--0.08\,M$_\odot$ \citep{Burgasser07},  the M-dwarf binary fraction is $f_{\rm bin} = 0.34$ for the primary mass range 0.08--0.47\,M$_\odot$ \citep{Janson12}, the K-dwarf binary fraction is $f_{\rm bin} = 0.45$ for the primary mass range 0.47--0.84\,M$_\odot$ \citep{Mayor92}, the G-dwarf binary fraction is $f_{\rm bin} = 0.46$ for the primary mass range 0.84--1.20\,M$_\odot$ \citep{Raghavan10}, the A-star binary fraction  is $f_{\rm bin} = 0.48$ for the primary mass range 1.20--3.00\,M$_\odot$ \citep{DeRosa14} and for more massive stars ($>3.00$\,M$_\odot$) the binary fraction is set to unity \citep[e.g.][]{Sana13}.

If a system is a binary, we define the mass ratio of the system, $q$, in terms of the primary mass $m_p$ and the secondary mass $m_s$ (the primary mass is always larger than or equal to the secondary mass):
\begin{equation}
q = \frac{m_s}{m_p},
\end{equation}
where $m_p$ is the mass drawn from the IMF. In the majority of simulations we assign secondary masses from a flat mass ratio distribution \citep{Reggiani11a,Reggiani13}, i.e.\,\,the secondary mass $m_s$ is determined by multiplying the primary mass $m_p$ by a random number between zero and unity. Occasionally, this produces a very low (planetary) mass secondary component, but usually the secondaries are stellar or brown dwarf objects.  In the set of simulations taken from \citet{DaffernPowell22}, we assign every companion to be Jupiter-mass ($m_s = 1$\,M$_{\rm Jup} = 9.5 \times 10^{-4}$\,M$_\odot$).

In the majority of simulations the binary semimajor axis distribution is a delta function, with all the binaries at 1, 30 or 200\,au, respectively, in the different simulations. In the simulations which adopt a version of the \citet{Kroupa95a} `Universal' binary population, the semimajor axes for all systems (irrespective of primary mass) are drawn from the following distribution,
\begin{equation}
f\left({\rm log_{10}}a\right) = \eta\frac{{\rm log_{10}} a - {\rm log_{10}} a_{\rm min}}{\delta + \left({\rm log_{10}} a - {\rm log_{10}} a_{\rm min}\right)^2},
\label{coma}
\end{equation}
where ${\rm log_{10}} a$ is the logarithm of the semi-major axis in au and ${\rm log_{10}} a_{\rm min} = -2$ ($a_{\rm min} = 0.01$\,au). The numerical constants are $\eta = 5.25$ and $\delta = 77$.

In the simulations that have a Field-like population of binaries, the semimajor axis distribution is a function of the primary mass. For systems with primary masses higher than 3.00\,M$_\odot$ (i.e. O- and B-type stars), the semimajor axis distribution is a log-flat \citet{Opik24} distribution, with semimajor axes between  0 -- 50\,au.

All other systems in these Field-like populations have semimajor axes drawn from a log-normal distribution, but with different mean and variance depending on the primary mass. For systems where the primary mass is a brown dwarf (0.01 -- 0.08\,M$_\odot$), the log-normal parameters are  ${\rm log_{10}} \bar{a} = 0.66$ (mean $\bar{a} = 4.6$\,au) and variance $\sigma_{{\rm log}_{10} \bar{a}} = 0.40$ \citep{Thies07}. For systems where the primary mass is an M-dwarf (0.08 -- 0.45\,M$_\odot$), the log-normal parameters are ${\rm log_{10}} \bar{a} = 1.20$ (mean $\bar{a} = 16$\,au) and variance $\sigma_{{\rm log}_{10} \bar{a}} = 0.80$ \citep{Janson12}. For systems where the primary mass is a G-dwarf (0.84 -- 1.2\,M$_\odot$) the log-normal parameters are ${\rm log_{10}} \bar{a} = 1.70$ (mean $\bar{a} = 50$\,au) and variance $\sigma_{{\rm log}_{10} \bar{a}} = 1.68$ \citep{Raghavan10}. For systems where the primary mass is an A-star (1.5--3.0\,M$_\odot$) the log-normal parameters are ${\rm log_{10}} \bar{a} = 1.70$ (mean $\bar{a} = 389$\,au) and variance $\sigma_{{\rm log}_{10} \bar{a}} = 0.79$.

All other primary mass ranges are assigned the log-normal paramaters for the observed G-dwarf distribution \citep{Raghavan10}.

In the simulations with Jupiter-mass companions from \citet{DaffernPowell22} the orbital eccentricites are set to zero (i.e.\,\,the orbits are perfectly circular). In the other simulations, the binary orbital eccentricities are drawn from a flat distribution \citep{Raghavan10}, apart from close ($a < 1$\,au) binaries which are thought to tidally circularise on short timescales \citep{Zahn77,Zahn89b} and therefore are assigned eccentricities of zero. 

\subsection{Star-forming regions}

To test the effects of subtle variations in the spatial and kinematic substructure of star-forming regions, we set up our simulated regions to be either very spatially and kinematically substructured, or to have none, or very little, substrcture.

For consistency, we model both of these scenarios using the box fractal method, which generates fractal substructure on all scales \citep{Goodwin04a}. We follow the method of \citet{Goodwin04a}, and start by defining a cube with sides of length $N_{\rm div} = 2$, within which we generate the star-forming region. A `parent' particle is placed at the centre of this cube, and then we divide the cube into $N_{\rm div}^3$ sub-cubes, with a `child' particle placed at the centre of each sub-cube.

The probability of a child particle becoming a parent itself is calculated according to $N_{\rm div}^{D-3}$, where $D$ is the required fractal dimension. Children that do not become parent particles are removed, along with all of their parent particles.  The children  who do become parents have a small amount of noise added to their position vector, to prevent the structure from having a gridded appearance.

Each child's sub-cube is itself divided into $N_{\rm div}^3$, and the process is repated until there is a generation with significantly more particles than the required number of stars for the simulation. Any remaining parent particles are then removed, so that only the final generation is left.

Finally, the region is pruned so that the particles sit within the boundary of a sphere, and if there are still more particles than the required number of stars, particles are removed at random until the desired number of stars is reached. This maintains the chosen fractal dimension as closely as possible.

The mean number of child particles that themselves become parents is $N_{\rm div}^D$, so that a lower fractal dimension corresponds to fewer particles maturing and becoming parent particles themselves. This results in a more substructured spatial distribution, whereas for a high fractal dimension (e.g. $D = 3.0$) nearly all the particles mature, and so the fractal has an almost uniform spatial distribution.

The fractals are kinematically substructured, such that stars that are physically closer together have a low velocity dispersion, whereas stars that are further apart have a higher dispersion, similar to the observed \citet{Larson81} relations.

\begin{figure}
\begin{center}
\rotatebox{270}{\includegraphics[scale=0.38]{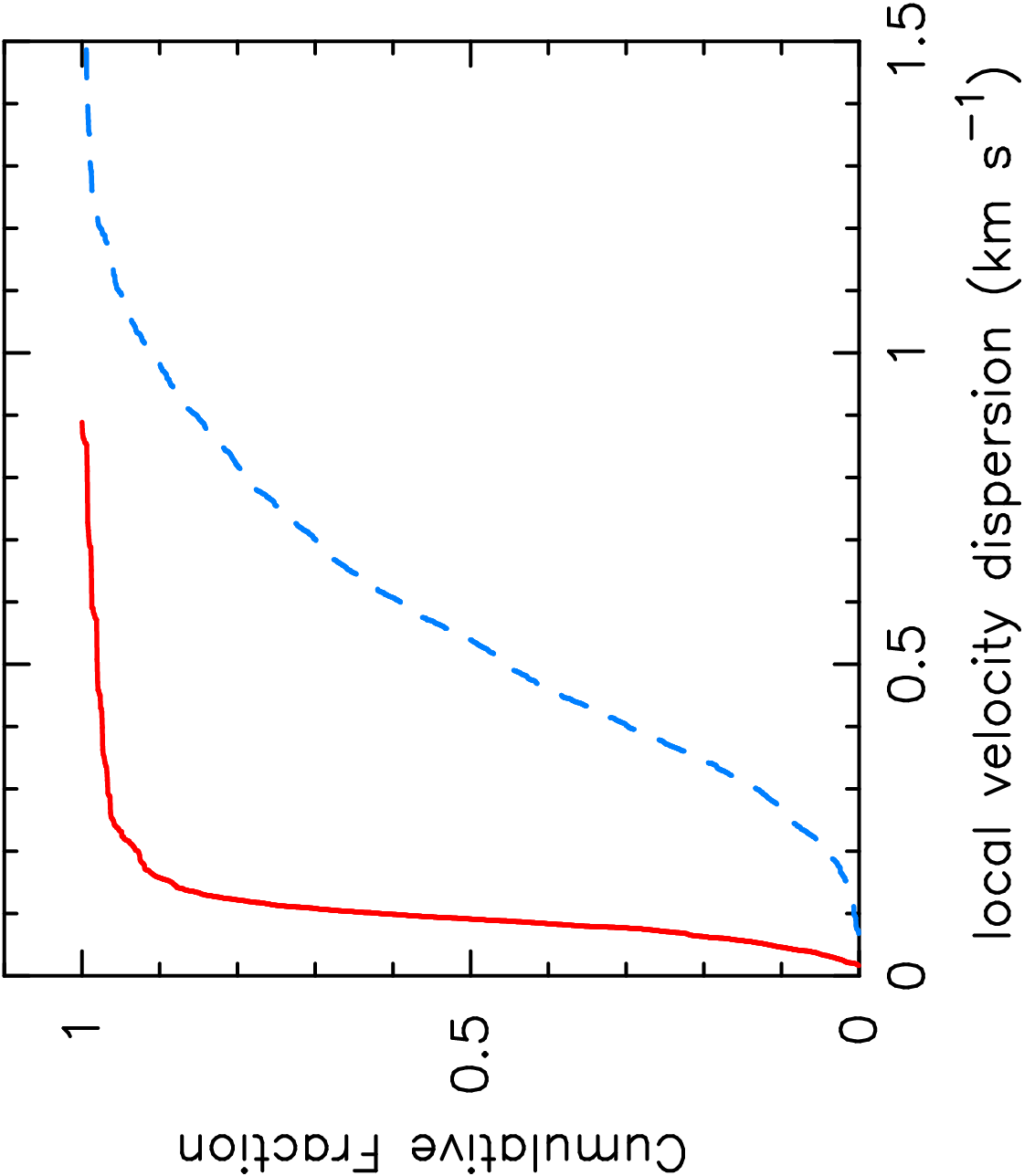}}
\caption[bf]{The initial distributions of local velocity dispersions in our simulations, after scaling to a subvirial virial ratio ($\alpha_{\rm vir} = 0.3$). For each star, we have calculated the velocity dispersion of the ten nearest neighbours around that star. The solid red curved line shows this distribution for the kinematically substructured initial conditions ($D = 1.6$) and the dashed blue curved line shows the distribution for non-kinematically substructured initial conditions ($D = 3.0$).}
\label{velocities}
  \end{center}
\end{figure}

The velocities in the fractal are set according to the method in \citet{Goodwin04a}. First, each parent particle has its velocity drawn from a Gaussian of mean zero. The stars that are placed in the final postions in the fractal inherit their parent particle's velocity, plus an additional random component drawn from the same Gaussian and multiplied by a factor $\left(\frac{1}{N_{\rm div}}\right)^g$, where $g$ is the integer number that describes the generation of particle in the fractal the star belongs to. This leads to the additional random components being smaller on average with each sucessive generation.

We scale the velocities so that the region has a virial ratio $\alpha_{\rm vir} = 0.3$ (which is a subvirial velocity distribution; $\alpha_{\rm vir} = 0.5$ corresponds to virial equilibrium, and is the ratio of total kinetic and potential energies of the stars).

In three dimensions, the largest different between two box fractal distributions is where one is very substructured with a fractal dimension $D = 1.6$ and the other extreme is a uniform sphere, with a fractal dimension $D = 3.0$.

The differences in the velocity dispersions between these regions are shown in Fig.~\ref{velocities}. We show the distribution of the local velocity dispersion (i.e.\,\,the velocity dispersion out to the tenth nearest neighbour for each star), where the velocities have been scaled to the overall virial ratio ($\alpha_{\rm vir} = 0.3$). Clearly, the velocity dispersions are much smaller in the highly fractal ($D = 1.6$) regions (the solid red curved line) compared to the smooth/uniform ($D = 3.0$) distributions (the dashed blue curved line).

We set the radii of the fractals such that the median local stellar densities (the median in a distribution in which we calculate the stellar volume density for each star out to its tenth nearest neighbour) are the same. For the more substructured $D = 1.6$ fractals, this radius is 1\,pc, whereas for the more uniform $D = 3.0$ fractals the radius is 0.25\,pc.

We summarise the different simulations, and their respective binary populations, in Table~\ref{simulations}.

\begin{table*}
  \caption[bf]{A summary of the different initial conditions of our simulated star-forming regions. The columns show the fractal dimension (amount of substructure) $D$, radius of the fractal $r_F$, median local stellar density $\tilde{\rho}$, the lower limit to the IMF $m_l$, the binary semimajor axis distribution $f(a)$, the binary mass ratio distribution, $f(q)$, and the binary eccentricity distribution $f(e)$. }
  \begin{center}
    \begin{tabular}{|c|c|c|c|c|c|c|}
      \hline

$D$ & $r_F$ & $\tilde{\rho}$ & $m_l$ & $f(a)$ & $f(q)$ & $f(e)$  \\
      \hline
      1.6 & 1\,pc & $10^4$\,M$_\odot$\,pc$^{-3}$ & 0.08\,M$_\odot$ & $a = 30$\,au & $m_s = 1$\,M$_{\rm Jup}$ & $e = 0$ \\
      3.0 & 0.25\,pc & $10^4$\,M$_\odot$\,pc$^{-3}$ & 0.08\,M$_\odot$ & $a = 30$\,au & $m_s = 1$\,M$_{\rm Jup}$ & $e = 0$ \\
      \hline
      1.6 & 1\,pc & $10^4$\,M$_\odot$\,pc$^{-3}$ & 0.01\,M$_\odot$ & $a = 30$\,au & Flat $q$  & Flat $e$ \\
      3.0 & 0.25\,pc & $10^4$\,M$_\odot$\,pc$^{-3}$ & 0.01\,M$_\odot$ & $a = 30$\,au & Flat $q$ & Flat $e$ \\
      \hline
      1.6 & 1\,pc & $10^4$\,M$_\odot$\,pc$^{-3}$ & 0.01\,M$_\odot$ & $a = 1$\,au & Flat $q$  & Flat $e$ \\
      3.0 & 0.25\,pc & $10^4$\,M$_\odot$\,pc$^{-3}$ & 0.01\,M$_\odot$ & $a = 1$\,au & Flat $q$ & Flat $e$ \\
      \hline
            1.6 & 1\,pc & $10^4$\,M$_\odot$\,pc$^{-3}$ & 0.01\,M$_\odot$ & $a = 200$\,au & Flat $q$  & Flat $e$ \\
      3.0 & 0.25\,pc & $10^4$\,M$_\odot$\,pc$^{-3}$ & 0.01\,M$_\odot$ & $a = 200$\,au & Flat $q$ & Flat $e$ \\
      \hline
            1.6 & 1\,pc & $10^4$\,M$_\odot$\,pc$^{-3}$ & 0.01\,M$_\odot$ & Field-like & Flat $q$  & Flat $e$ if $a \geq 1$\,au; $e = 0$ if $a<1$\,au  \\
      3.0 & 0.25\,pc & $10^4$\,M$_\odot$\,pc$^{-3}$ & 0.01\,M$_\odot$ & Field-like & Flat $q$ & Flat $e$ if $a \geq 1$\,au; $e = 0$ if $a<1$\,au \\
      \hline
            1.6 & 1\,pc & $10^4$\,M$_\odot$\,pc$^{-3}$ & 0.01\,M$_\odot$ & \citet{Kroupa95a} `Universal' (Eqn.~\ref{coma}) & Flat $q$  & Flat $e$ if $a \geq 1$\,au; $e = 0$ if $a<1$\,au  \\
      3.0 & 0.25\,pc & $10^4$\,M$_\odot$\,pc$^{-3}$ & 0.01\,M$_\odot$ & \citet{Kroupa95a} `Universal' (Eqn.~\ref{coma}) & Flat $q$ & Flat $e$ if $a \geq 1$\,au; $e = 0$ if $a<1$\,au \\
      \hline

    \end{tabular}
  \end{center}
  \label{simulations}
\end{table*}

\subsection{Dynamical evolution and identification of bound binaries}

We randomly place the stellar systems (either single or binary, where binaries can have stellar or planetary companions) at the positions of the `stars' in the fractals. We use the $4^{\rm th}$-order Hermite $N$-body integrator \texttt{kira}  within the \texttt{Starlab} environment \citep{Zwart99,Zwart01} to evolve the star-forming regions for 10\,Myr. We do not include stellar evolution in the simulations.

Throughout the simulations, a pair of stars are considered a bound binary if they are mutual nearest neighbours, \emph{and} they have a total negative energy, $E_{\rm bind}$, which we can write in terms of the semimajor axis $a$ and  masses of the primary and secondary componments, $m_p$ and $m_s$: 
\begin{equation}
  E_{\rm bind} = -\frac{Gm_pm_s}{2a}.
  \label{binding_energy}
\end{equation}
Due to the high stellar density of our simulations, not all of the primordial binaries we place into the simulations fulfill the criteria of being mutual nearest neighbours and energetically bound.

Finally, we note that systems with a high binding energy have smaller semimajor axes (i.e.,\,\,they are closer), and these systems also have higher orbital velocities. Depending on the stellar environment, there is a boundary between `hard' (fast, close) systems and `soft' (slow, wide) systems. Usually, hard systems become harder (faster/closer) following an interaction(s) and soft systems become softer (slower/wider) -- this is the Heggie-Hills law \citep{Heggie75,Hills75a,Hills75b}.

\section{Results}
\label{results}

\subsection{Delta function 30\,au}

\begin{figure*}
  \begin{center}
\setlength{\subfigcapskip}{10pt}
\hspace*{-1.5cm}\subfigure[Companions all 1\,M$_{\rm Jup}$]{\label{30au_comp-a}\rotatebox{270}{\includegraphics[scale=0.37]{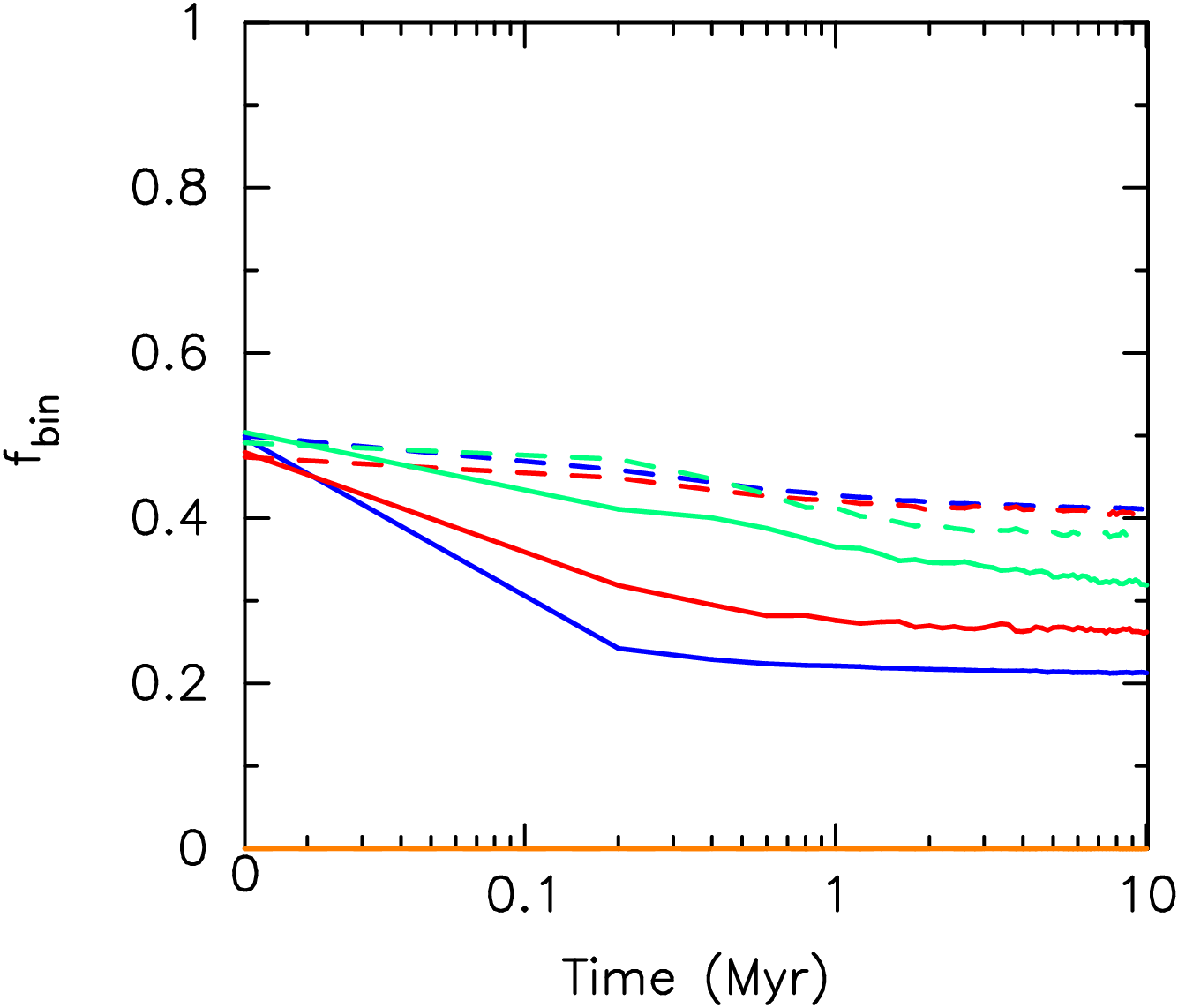}}}
\hspace*{0.3cm} 
\subfigure[Companions from flat $q$]{\label{30au_comp-b}\rotatebox{270}{\includegraphics[scale=0.37]{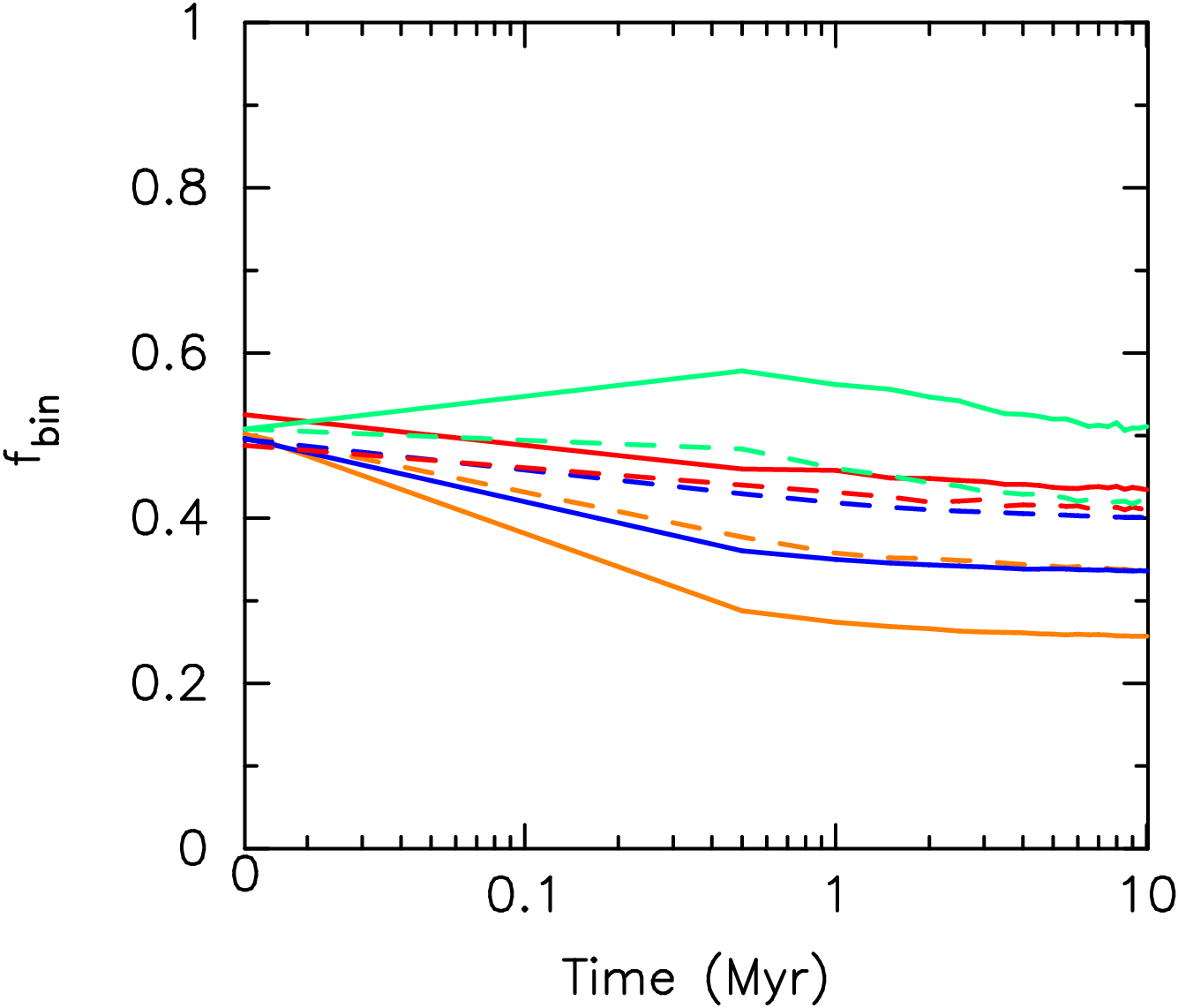}}}
\caption[bf]{Evolution of the binary fraction when companions are all initially placed at 30\,au from the host star. In panel (a) we show the results when all the companions are planetary-mass (1\,M$_{\rm Jup}$) and in panel (b) we show the results when the companions are drawn from a flat mass ratio distribution. In both panels, the solid  curved lines represent substructured simulations ($D = 1.6$) and the dashed  curved lines represent non-substructured simulations ($D = 3.0$). The green curved lines are for binaries with A-type primary stars, the red  curved lines are for binaries with G-type primary stars, the blue curved lines are for binaries with M-type primary stars and the orange lines are for binaries with brown dwarf primaries (the simulations shown in panel (a) do not contain brown dwarf primaries).   }
\label{30au_comp}
  \end{center}
\end{figure*}

In Fig.~\ref{30au_comp-a} we show the evolution of the binary fraction from the simulations in \citet{DaffernPowell22}, in which a single Jupiter-mass planet was placed on orbit at 30\,au around 50\,per cent of the stars, and in Fig.~\ref{30au_comp-b} simulations where companion masses are drawn from a flat mass ratio distribution and then placed on 30\,au orbits.

The coloured lines show the evolution of the binary fractions for different primary masses. Green lines are for systems where the primary is an A-type star ($1.5 < m/{\rm M_\odot} \leq 3.0$), red lines are systems where the primary is a G-type star ($0.8 < m/{\rm M_\odot} \leq 1.2$), blue lines are systems where the primary is a M-type star ($0.08 < m/{\rm M_\odot} \leq 0.5$) and orange lines are systems where the primary is a brown dwarf (these systems were not included in the simulations shown in panel (a)).

The solid lines represent the simulations with a high degree of initial spatial and kinematic substructure (where the fractal dimension $D = 1.6$). The dashed lines represent simulations with no spatial and kinematic substructure (where the fractal dimension $D = 3.0$).

The significantly lower binary fraction for the substructured simulations (the solid lines) reflects the results from \citet{DaffernPowell22}, who found that significantly more planets become unbound and free-floating in substructured simulations compared to non-substructured simulations.

However, there is a strong dependence of destruction on the system mass; in the simulations where all the companions are planetary mass, more systems are destroyed than when the companions are drawn from a flat mass ratio distribution (compare panel (a) with panel (b)). Note that it is the overall system mass that determines the likelihood of destruction, not the companion mass ratio \citep{Goodwin13,Parker13b}.

In the simulations where the companions are drawn from a flat mass ratio distribution (Fig.~\ref{30au_comp-b}) the binary fraction for G- and A-type stars is higher in the substructured simulations. This is because a significant number of systems form via dynamical capture \citep{Kouwenhoven10,Moeckel10}, especially if the velocities are initially correlated, as they are in the substructured simulations. Furthermore, it is easier to create dynamical binaries if the overall system mass is high, which is why very few M-dwarf binaries form through this mechanism. We performed a check where we repeated the calculation but excluded systems that formed via capture and found that the binary fractions are all lower for the kinematically substructured simulations.

\begin{figure}
\begin{center}
\rotatebox{270}{\includegraphics[scale=0.38]{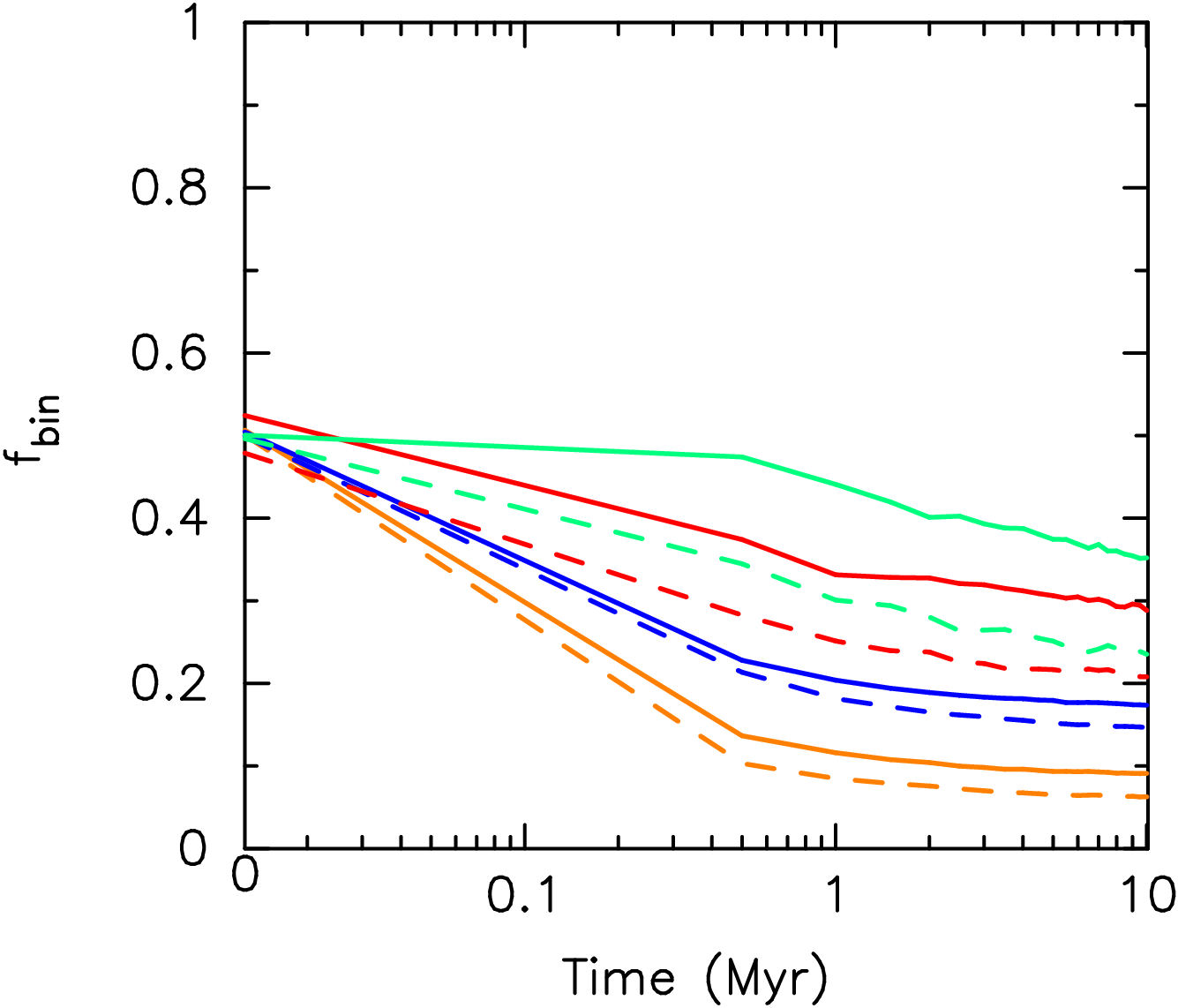}}
\caption[bf]{Evolution of the binary fraction when companions are all initially placed at 200\,au from the host star and have a flat mass ratio distribution. The solid  curved lines represent substructured simulations ($D = 1.6$) and the dashed  curved lines represent non-substructured simulations ($D = 3.0$). The green  curved lines are for binaries with A-type primary stars, the red curved lines are for binaries with G-type primary stars, the blue  curved lines are for binaries with M-type primary stars and the orange curved lines are for binaries with brown dwarf primaries.}
\label{fbin_200au_delta}
  \end{center}
\end{figure}

\subsection{Delta function 1\,au}

We also run a set of simulations in which the binary companions were placed at 1\,au from the host star. These binaries have significantly higher binding energies than those with companions at 30\,au (i.e.\,\,they are harder/faster). We see the same behaviour as for the 30\,au binaries, whereby the systems in the kinematically substructured star-forming regions ($D = 1.6$) are more readily destroyed than the systems in the non-substructured star-forming regions ($D = 3.0$).

\subsection{Delta function 200\,au}

We now place the binary companions at 200\,au and draw companion masses from a flat distribution. The results are shown in Fig.~\ref{fbin_200au_delta}. In this plot, we see that the binary fraction for brown dwarf, and M-dwarf primaries is lower for the non-substructured simulations ($D = 3.0$), compared to the substructured simulations ($D = 1.6$), i.e.\,\,opposite to the case where all companions are placed at 30\,au or 1\,au. The same behaviour is seen for G- and A-type primaries, although as for the 30\,au binaries this difference is slightly exacerbated by the  numbers of systems witha G- or A-type primary that form via capture in the substructured simulations.


\subsection{Field distribution}

We now discuss simulations in which we place the field binary population as the initial binary population. As discussed in Section~\ref{methods}, the field population is characterised by a decrease in the mean semimajor axis with decreasing primary mass (A-type stars peak at $\sim 390$\,au \citep{DeRosa14}, G-type binaries peak at $\sim 50$\,au \citep{Raghavan10}, M-type primaries peak at $\sim 16$\,au \citep{Ward-Duong15} and brown dwarf binaries peak at $\sim 5$\,au \citep{Burgasser07}). This means that we might expect the majority of A-type binaries (and to a lesser extent, G-type binaries) to be dynamically softer/slower than the M-type and brown dwarf binaries, despite their higher system masses.

\begin{figure}
\begin{center}
\rotatebox{270}{\includegraphics[scale=0.38]{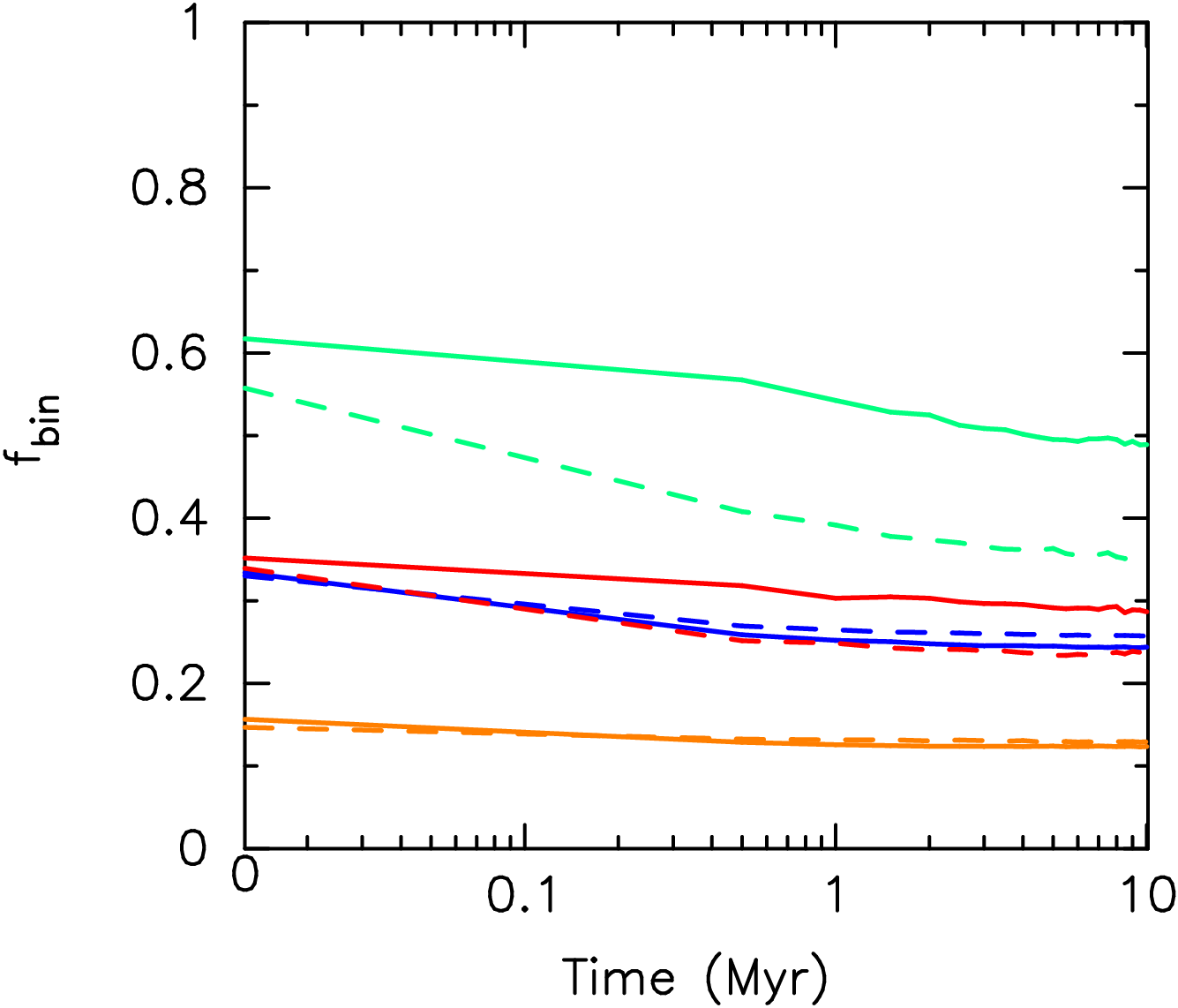}}
\caption[bf]{Evolution of the binary fraction when companions are assigned based on the observed semimajor axis distributions in the Galactic field, and have a flat mass ratio distribution. The solid  curved lines represent substructured simulations ($D = 1.6$) and the dashed  curved lines represent non-substructured simulations ($D = 3.0$). The green curved lines are for binaries with A-type primary stars, the red lines are for binaries with G-type primary stars, the blue curved lines are for binaries with M-type primary stars and the orange  curved lines are for binaries with brown dwarf primaries. }
\label{fbin_field}
  \end{center}
\end{figure}

The evolution of the binary fraction for this population is shown in Fig.~\ref{fbin_field}. Here, the substructured simulations destroy more brown dwarf and M-dwarf systems than the smooth simulations (although the differences are of order of only a few per cent), whereas more G-type and A-type stars form binaries via capture in the substructured simulations and so the binary fractions are higher for A- and G-type stars than in the non-substructured simulations.

These trends are broadly similar to the simulations where all companions were placed at 30\,au or 1\,au, and is due to the Field distributions being dominated by M-dwarf primary systems, whose binaries typically have semimajor axes less than 30\,au.

\subsection{\citet{Kroupa95a} distribution}

Finally, we look at the evolution of binary fractions in kinematically substructured simulations, and non-substructured simulations, where the binary population is drawn from the \citet{Kroupa95a} distribution, in which all the binaries have the same semimajor axis distribution, and the binary fraction is nominally unity.

\begin{figure}
\begin{center}
\rotatebox{270}{\includegraphics[scale=0.38]{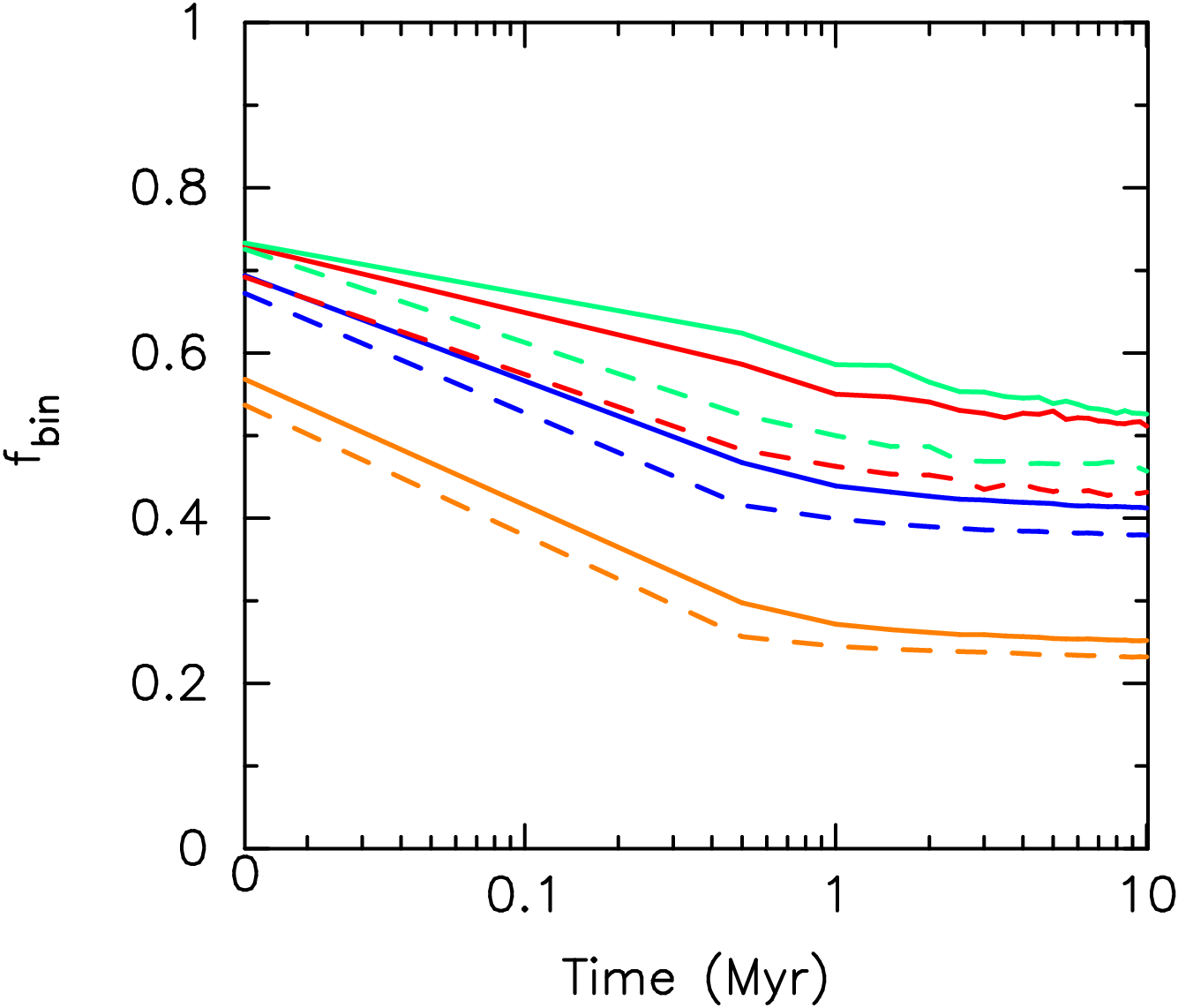}}
\caption[bf]{Evolution of the binary fraction when companions are drawn from a `universal' semimajor axis distribution \citep{Kroupa95a}, and have a flat mass ratio distribution. The solid curved lines represent substructured simulations ($D = 1.6$) and the dashed  curved lines represent non-substructured simulations ($D = 3.0$). The green  curved lines are for binaries with A-type primary stars, the red curved lines are for binaries with G-type primary stars, the blue curved lines are for binaries with M-type primary stars and the orange curved  lines are for binaries with brown dwarf primaries.}
\label{fbin_kroupa}
  \end{center}
\end{figure}

The \citet{Kroupa95a} distribution contains an overabundance of wide ($>$1000\,au) binaries. However, as the simulations are very dense, many of the widest pairs will not be physically bound, despite being placed in the simulation as primordially bound systems. For this reason, the initial binary fraction (the coloured lines at $t = 0$\,Myr in Fig.~\ref{fbin_kroupa}) are not at unity, but instead lie between $\sim$0.55 for the brown dwarfs, to $\sim$0.73 for the A-type stars (more massive systems tend to have higher binding energies, and so more A-type binaries are bound).

However, in the substructured simulations the initial binary fractions are already slightly higher than in the smooth simulations because the kinematic substructure creates more wide but bound pairs (compared to the more random velocities in the smooth simulations).

At all stages in the simulations the binary fraction in the substructured simulations is higher than in the smooth simulations, and the G-type and A-type binary fractions decrease faster in the smooth simulations than in the substructured simulations. We might expect this to be the case given that this distribution is dominated by systems with much wider semimajor axes than in the simulations where all binaries are placed at 200\,au. 

\section{Discussion}
\label{discuss}

Our results show that for binary populations with significant numbers of intermediate to small-separation systems ($<$100\,au), such as the Galactic field population, more binary systems are destroyed in spatially and kinematically substructured star-forming regions, compared to smooth regions of similar density. However, when the binary populations are dominated by wider systems ($>$100\,au), more systems are destroyed in smooth regions, comapred to substructured regions (again, with identical local densities).

Our explanation for this apparently contradictory behaviour comes from considering the hard-soft boundary for binary destruction \citep{Heggie75,Hills75a,Hills75b}, also referred to as the fast-slow boundary \citep{Fregeau04,Fregeau06}.

The typical energy of a star in a star-forming region (sometimes referred to as the `Maxwellian energy') is approximated by assuming an average mass $\langle m \rangle$, and a typical velocity dispersion within the region, $\sigma$: 
\begin{equation}
E_{\rm Max} \approx  \langle m \rangle \sigma^2.
\end{equation}
For a binary to be classed as dynamically `hard', i.e. unlikely to be destroyed by an encounter, it has to have a binding energy $E_{\rm bind}$, in excess of the local Maxwellian energy, such that it satisfies
\begin{equation}
|E_{\rm bind}| > \langle m \rangle \sigma^2.
\end{equation}
If we re-cast this in terms of the orbital velocity of the binary, $v_{\rm orb} > \sigma$. Conversely if a binary is dynamically `soft', i.e.\,\,likely to be destroyed, then its binding energy will be less than the local Maxwellian energy, such that
\begin{equation}
  |E_{\rm bind}| < \langle m \rangle \sigma^2.
\end{equation}
Again, in terms of the orbital velocity of the binary, such a system would be `slow', and its orbital velocity would be less than the local velocity dispersion, $v_{\rm orb} < \sigma$.

Recall that the binding energy of a binary is inversely proportional to the semimajor axis (Eqn.~\ref{binding_energy}), and in the following argument we will assume that the component masses ($m_p, m_s$) of the binary are constant. We can therefore re-write Eqn~\ref{binding_energy} as
  \begin{equation}
|E_{\rm bind}| = \frac{C}{a},
  \end{equation}
  where the constant $C = Gm_pm_s/2$. 
  We then consider how the binding energy relates to the local Maxwellian energy in the cluster for hard (fast) and soft (slow) binaries).

In the case of hard (fast) binaries, then
\begin{equation}
  \frac{C}{a} > \sigma^2
\end{equation}
so for small $a$ ($<100$\,au), more binaries are destroyed if $\sigma$ is (relatively) small, which is the case in the highly substructured ($D = 1.6$, the solid red line in Fig.~\ref{velocities}) simulations.

In the case of soft (slow) binaries, then
\begin{equation}
\frac{C}{a} < \sigma^2
\end{equation}
so for large $a$ ($>100$\,au), more binaries are destroyed if $\sigma$ is (relatively) large, which is the case for the smooth, non-substructured ($D = 3.0$, the dashed blue line in Fig.~\ref{velocities}) simulations.

This transition at the fast--slow boundary for the efficacy of binary destruction in substructured star-forming regions would be unlikely to be observed in reality. If star-forming regions all start with the same (`Universal') population of binaries \citep{Kroupa95a}, in order to compare two regions we would need to establish that the initial local densities were similar.

Whilst it is possible to infer the initial conditions of a star-forming region using measures of substructure, mass segregation and the velocities of ejected stars, even combinations of these techniques can only pinpoint initial stellar densities to within a factor of two at best, and such a difference in stellar densities could also be responsible for the $\sim$5\,per cent difference in binary fractions in our simulations (e.g. Fig.~\ref{fbin_kroupa}).

A more promising avenue for further research is in the host stars of gas giant planets. The formation locations of gas giants in protoplanetary discs are likely to be well within 100\,au, due to an absence significant numbers of discs with gas radii in excess of this \citep[e.g.][]{Drazkowska22}. Therefore, we would expect planet host stars that formed in very substructured environments to be more likely to lose their planets, than stars that formed in smooth star-forming regions, and as we have seen in Fig.~\ref{30au_comp-a}, there could be a $>20$\,per cent difference in the bound giant planet fraction.

When the binary components are more massive ($m_p \geq 1$\,M$_\odot$ and $m_s \geq 0.01$\,M$_\odot$, rather than $m_s = 9.4 \times 10^{-4}$\,M$_\odot$) a significant number of binaries form via capture \citep{Kouwenhoven10,Moeckel10,Parker14d}, which alters the binary fractions for the more massive systems. We have included these systems in our analysis, because observationally it would be extremely difficult to distinguish primordial binary systems from those that formed via capture. However, these systems also add to the confusion in interpreting the differences in binary fractions between substructured and smooth star-forming regions.

\section{Conclusions}
\label{conclude}

We present $N$-body simulations of the destruction of binary systems in star-forming regions with and without spatial and kinematic substructure, but where we have kept the local stellar density constant for each type of simulation, to determine the effect of the substructure on binary processing. Our conclusions are the following:

(i) In simulations where the binary population is dominated by close ($<$100\,au) binaries (e.g. the Galactic field population, or a delta function at close separations), the substructured simulations destroy more binary systems.

(ii) In simulations where the binary population is dominated by wider ($>$100\,au) binaries (e.g. the \citet{Kroupa95a} `Universal' population, or a delta function at wide separations), the non-substructured (smooth) simulations detroy more binary systems.

(iii) Our interpretation of this is that the hard (close), or fast binaries are more readily destroyed when the local velocity dispersion is small (i.e.\,\,velocities are correlated on local scales, as is the case for the substructured simulations). Conversely, the soft (wide), or slow, binaries are more readily destroyed when the local velocity dispersion is large (velocities are not correlated on local scales, as is the case for the non-substructured simulations).

(iv) For stellar and brown dwarf binaries, the difference in overall binary fraction between the substructured and non-substructured simulations is around 5\,per cent, which is usually within the observational uncertainties of the binary fraction in star-forming regions \citep{Duchene13b}.

(v) However, for planetary systems, the initial semimajor axis range is much smaller than for stellar binaries, and these systems exhibit significantly different binary fractions between the substructured and non-substructured simulations. We therefore suggest that planetary populations that are affected by stellar flybys may exhibit subtle differences due to the substructure (or lack thereof) in their birth star-forming region.

\section*{Acknowledgements}

I dedicate this paper to the memory of my friend, Alan~Chopping, who sadly passed away in April 2023. Before he retired, Alan was an operations engineer at the Isaac Newton Group of telescopes, helping generations of staff, students and visiting astronomers obtain data. \\ 

\noindent I acknowledge support from the Royal Society in the form of a Dorothy Hodgkin Fellowship. I thank the anonymous referee for a helpful report.

\section*{Data availability statement}

The data used to produce the plots in this paper will be shared on reasonable request to the corresponding author.

\bibliographystyle{mnras}  
\bibliography{general_ref}

\appendix

\label{lastpage}

\end{document}